\documentclass[conference,a4paper]{IEEEtran}
\IEEEoverridecommandlockouts

\usepackage{cite}
\usepackage{amsmath,amssymb,amsfonts}
\usepackage{algorithmic}
\usepackage{graphicx}
\usepackage{textcomp}
\usepackage{xcolor}
\usepackage{colortbl}
\definecolor{headblue}{HTML}{DCEAF7}
\definecolor{useblue}{HTML}{EAF3FB}
\definecolor{strengthgreen}{HTML}{E7F4E4}
\definecolor{problemred}{HTML}{FBE7E4}
\definecolor{edgeamber}{HTML}{FFF2CC}
\definecolor{decisiongray}{HTML}{EEEEEE}
\usepackage{xurl}
\def\BibTeX{{\rm B\kern-.05em{\sc i\kern-.025em b}\kern-.08em
    T\kern-.1667em\lower.7ex\hbox{E}\kern-.125emX}}

\usepackage{multirow}
\usepackage{tikz}
\usetikzlibrary{shapes.geometric, arrows.meta, positioning, fit, backgrounds, calc}
\makeatletter
\newcommand{\linebreakand}{%
  \end{@IEEEauthorhalign}
  \hfill\mbox{}\par
  \mbox{}\hfill\begin{@IEEEauthorhalign}
}
\makeatother

\begin{document}
\title{Reverse Engineering Compliance: A Dual-Graph Verification Framework for Auditing Legacy IT Security Concepts}
\author{%
\IEEEauthorblockN{Lea Muth}
\IEEEauthorblockA{Department of Mathematics and Computer Science \\
Freie Universität Berlin\\
Berlin, Germany \\
Lea.Muth@fu-berlin.de*}
\and
\IEEEauthorblockN{Marian Margraf}
\IEEEauthorblockA{Department of Mathematics and Computer Science \\
Freie Universität Berlin\\
Berlin, Germany \\
Marian.Margraf@fu-berlin.de}
}

\maketitle

\begingroup
\renewcommand\thefootnote{}
\footnotetext{\textcopyright~2026 IEEE. Personal use of this material is permitted. Permission from IEEE must be obtained for all other uses, in any current or future media, including reprinting/republishing this material for advertising or promotional purposes, creating new collective works, for resale or redistribution to servers or lists, or reuse of any copyrighted component of this work in other works.}
\endgroup
 
\begin{abstract}
The NIS-2 Directive increases the need for continuous, auditable compliance evidence and motivates a shift from document-based compliance toward machine-readable compliance artifacts. The Open Security Controls Assessment Language (OSCAL) is a standard for this purpose, which the German Federal Office for Information Security (BSI) is adapting with Grundschutz++. However, companies are still managing extensive legacy IT security concepts (IT-SCs), and migrating them without verification could transfer outdated assets into the new format. While existing research primarily addresses the generation of new concepts, there is a lack of a verification framework that extracts legacy IT-SCs into an auditable intermediate representation, deterministically compares the extracted graph with an independently constructed reference state, and exports schema-valid OSCAL artifacts.
This paper introduces the Automated Security Concept Structure Extraction and Reverse Topology-checking (ASSERT) Framework, which addresses this gap by using ontology-based extraction of legacy documents into formal document graphs, a five-class graph difference against a verified reference graph, and the export into schema-valid OSCAL outputs for system description and assessment evidence. Using the BSI's RecPlast dataset, we compare a local open-weight model and a commercial model across three configurations with different levels of reference-ontology exposure. The evaluation shows that ASSERT makes document-infrastructure inconsistencies measurable, but reveals a trade-off between discovering undocumented entities and enforcing a schema.
\end{abstract}


\section{Introduction}
With the German implementation of the European NIS-2 Directive~\cite{NIS2} effective since 6 December 2025, IT risk management is shifting from manual, document-based compliance toward automated, data-driven compliance (Compliance as Code (CaC)).
A key technical foundation for this shift is the Open Security Controls Assessment Language (OSCAL)~\cite{NISTOSCAL2026}, a machine-readable format developed by the National Institute of Standards and Technology (NIST). Structured data formats are not mandated by NIS-2 directly, but provide the technical foundation for continuous, scalable, and auditable compliance automation. In Germany, the Federal Office for Information Security (BSI) operationalizes this direction through Grundschutz++, which adopts OSCAL as a data model and moves the previously document-based IT-Grundschutz standard toward data-driven, evidence-based compliance artifacts.
Although OSCAL defines the target format, many companies remain bound to legacy IT Security Concepts (IT-SCs) that are outdated, contain copy-paste errors, omit newly added assets, and include expired security measures. An unverified translation of these IT-SCs into OSCAL would follow the ``garbage in, garbage out'' principle. During the multi-year transition between IT-Grundschutz and Grundschutz++ (beginning in January, 2026), German organizations must still account for legacy IT-SCs alongside emerging machine-readable artifacts.
Previous approaches toward automated compliance auditing mostly use Natural Language Processing (NLP) methods and Large Language Models (LLMs) to generate artifacts or prioritize vulnerabilities. Their probabilistic nature conflicts with the reproducibility and deterministic evidence handling required in security audits whenever an incorrect token is statistically more plausible than the factually correct one.
Our previous work~\cite{Paper2} addresses the generation of new compliant artifacts from an organization's raw data through the infrastructure graph and IT-SC (forward path) in IT-Grundschutz. The complementary migration problem remains unresolved, as an architecture that translates legacy IT-SCs into a formal intermediate representation that can be reproducibly verified against an external infrastructure reference graph (GT) is missing. This leads to the research question of this paper. How can inconsistencies between existing, potentially flawed legacy IT-SCs and the company's verified reference graph be identified, classified, and quantified? 
To address this question, our work proposes the \textbf{A}utomated \textbf{S}ecurity Concept \textbf{S}tructure \textbf{E}xtraction and \textbf{R}everse \textbf{T}opology-checking (ASSERT) Framework. 

ASSERT evaluates whether the topology constructed from the legacy IT-SC is consistent with a verified reference graph of the company infrastructure. The framework uses ontology-based information extraction to convert a legacy IT-SC into a document graph ($G_{Doc}$). For traceability, each extracted entity links to its original text paragraph. Instead of leaving content auditing to error-prone LLMs, $G_{Doc}$ is algorithmically compared with the company's verified reference graph ($G_{GT}$).
This deterministic graph comparison strictly decouples probabilistic text generation from verification logic and makes compliance errors measurable through the graph difference ($\Delta G$). The evaluation varies how much reference-ontology information is exposed during document-to-graph construction. For evaluation, node- and edge-level faults are injected into the reference IT-SC to assess the framework's detection capability.
The main contributions of this paper are:
\begin{enumerate}
\item \textbf{ASSERT Framework:} Ontology-based construction of an attributed document graph ($G_{Doc}$) from a legacy IT-SC, with complete traceability of each extracted entity back to its original passage.
\item \textbf{Formal error taxonomy:} Five disjoint, set-theoretic node- and edge-level error classes quantify compliance errors as a measurable graph difference $\Delta G$.
\item \textbf{Schema-valid OSCAL outputs:} Schema-valid NIST OSCAL v1.1.3 System Security Plan and Assessment Results artifacts.
\end{enumerate}
\section{Background}
\subsection{BSI IT-Grundschutz, Grundschutz++ and NIST OSCAL}
The BSI IT-Grundschutz methodology provides the foundation for the Information Security Management System in German government agencies and critical infrastructure (CI). The process requires a structural analysis (SA) of the IT environment, identification of protection needs for each asset, mapping of predefined requirements onto those assets, and, where necessary, risk analysis for assets without predefined requirements.
The result is the IT-SC, traditionally a monolithic text document in PDF format. Its text-based nature leads to operational inefficiencies, as the concepts rapidly become outdated, are prone to errors during manual updates, and cannot be audited in a scalable, automated manner. To address this issue, the BSI introduced ``Grundschutz++", whose core is OSCAL v1.1.3 developed by NIST. In OSCAL, the System Security Plan (SSP) describes the system, its boundaries, and control implementation, while Assessment Results (AR) record assessment actions, findings, and evidence mapped to those controls. Migrating legacy IT-SCs into these artifacts requires structural integrity and a justification of how each measure was implemented.
\subsection{Graph-Based Compliance Verification}\label{sec:background_graphs}
Converting a structured IT-SC into a machine-readable format requires an intermediate representation capable of capturing topological dependencies and semantic properties.
A company can be described as a directed, attributed graph $G=(V, E)$. The set of vertices $V$ represents entities such as business processes, IT systems, rooms, or applications. The set of edges $E$ defines the causal or topological relationships between these entities. 
There are two instances of this graph: the documented state $G_{Doc}$, constructed by ASSERT from the legacy IT-SC, and the verified reference state $G_{GT}$. The latter is derived from the company's operational structural data and forms a source of truth independent of the legacy IT-SC. We describe the construction of $G_{GT}$ from those sources using a SA agent that has been validated in our prior work~\cite{Paper2}. 
The advantage of this representation is the possibility of deterministic verification through graph matching. To compare the documented state ($G_{Doc} = (V_{Doc}, E_{Doc})$) with the reference state ($G_{GT} = (V_{GT}, E_{GT})$), we define the compliance difference $\Delta G$ as the 5-tuple:
\[
\Delta G = (V_O,\; V_P,\; E_O,\; E_T,\; E_G)
\]
The classification follows a cascading logic. First, structural matches and discrepancies are determined at the node level. Topological edge discrepancies are evaluated only after the matched-node set $V_{Matched}$ has been established. This ensures a disjoint error classification. Without restricting the edge level to $V_{Matched}$, a missing node $v \in V_O$ with $k$ incident edges $\{e_1, \ldots, e_k\} \subset E_{GT}$ would be recorded as $1 + k$ separate errors (once as node omission and $k$ times as edge omission), even though it is a single error. This is resolved by evaluating edges only between nodes in the matched set $V_{Matched}$. Edges connected to missing ($V_O$) or hallucinated ($V_P$) nodes are treated as a consequence of the node error, not as independent.
\newline
\textbf{Stage 1. Node Level (Existence Check):}
\newline
\textbf{Node Omissions ($V_O$):} Entities that exist in $G_{GT}$ but are not mentioned in the legacy IT-SC.
\[
V_O = V_{GT} \setminus V_{Doc}
\]
\newline
\textbf{Phantom Nodes ($V_P$):} Entities listed in the IT-SC but that do not exist in $G_{GT}$, such as obsolete or hallucinated assets.
\[
V_P = V_{Doc} \setminus V_{GT}
\]
Entities identified in both $G_{GT}$ and the IT-SC form the matched set:
\[
V_{Matched} = V_{GT} \cap V_{Doc}
\]
\textbf{Stage 2. Edge Level (Relation Check):}
\newline
\textbf{Edge Omissions ($E_O$):} Relationships between correctly identified nodes that exist in $G_{GT}$ but are missing in the IT-SC.
\[
E_O = \{(u,v) \in E_{GT} \mid u,v \in V_{Matched} \land (u,v) \notin E_{Doc}\}
\]
\newline
\textbf{Topological Conflicts ($E_T$):} Edges between correctly identified nodes that the IT-SC claims exist but do not exist in $G_{GT}$.
\[
E_T = \{(u,v) \in E_{Doc} \mid u,v \in V_{Matched} \land (u,v) \notin E_{GT}\}
\]
\newline
\textbf{Ghost Edges ($E_G$):} Edges connected to at least one phantom node ($V_P$). These edges are artifacts of hallucinated or obsolete entities and are not considered standalone topological errors, but are isolated as a consequence of the phantom nodes.
\[
E_G = \{(u,v) \in E_{Doc} \mid u \in V_P \lor v \in V_P\}
\]
\section{Related Work}
\subsection{The Missing Reverse Path in Compliance Automation}
AI-driven compliance automation has evolved from prompt engineering to structured hybrid architectures. \cite{Hsia2025} combines LLMs with Satisfiability Modulo Theories (SMT) solvers, in which the LLM translates regulatory requirements into formal SMT constraints and the solver deterministically verifies logical consistency. In production settings, ComplianceNLP~\cite{ComplianceNLP2026} reports a four-month parallel deployment with NER F1 of 91.3\%, gap-detection F1 of 87.7\%, and 70B-to-8B distillation with 98.6\% NER retention, showing that AI-assisted regulatory gap analysis has moved beyond the prototype stage. Further, Knowledge Graph-Augmented Retrieval-Augmented Generation (KG-RAG) has established itself as a dominant architectural pattern. Systems such as the ISO 27000 RAG framework~\cite{ISO27000RAG2026} and PrivComp-KG~\cite{PrivCompKG2024} use KGs to preserve the cross-references between standards that are critical for regulatory texts. In addition, compliance-specific benchmarks are emerging. COMPL-AI~\cite{COMPLAI2024} is the first to translate the requirements of the EU AI Act into measurable technical criteria for LLMs. Furthermore, the Bench-2-CoP study~\cite{Bench2CoP2025} quantifies a fundamental regulatory gap, as $0\%$ of the 194.955 benchmark questions analyzed cover the systemic risks required by the EU AI Act, such as self-replication or bypassing human supervision. Analyses of established frameworks~\cite{McIntosh2024} show that neither ISO 27001, ISO 42001, COBIT 2019, nor the NIS-2 Directive~\cite{NIS2} provide a comprehensive solution for managing LLM-specific risks. In particular, the risk of hallucinations is not systematically mitigated across these frameworks, highlighting the necessity of expert supervision.

Regardless of these advances, an asymmetry in compliance research is apparent. The literature focuses on forward compliance, that is, the development of new systems, the review of current guidelines, and the identification of new gaps in existing regulations. However, the industry holds a substantial amount of outdated IT-SCs, whose contents no longer correspond to the actual infrastructure. The reverse engineering of these inaccurate legacy IT-SCs into deterministic, verifiable topologies remains, to our knowledge, unexplored in the literature. Our previous study~\cite{Paper2} demonstrates this asymmetry, as LLM agents excel at information gathering in the forward-engineering approach but fail when applying deterministic inference rules. These findings are consistent with broader evidence that probabilistic LLMs struggle with formal reasoning and regulatory logic, especially when tasks require multi-step deductions over interdependent requirements~\cite{McIntosh2024, Hsia2025, Bench2CoP2025, Kambhampati2024}. Moreover, \cite{Valmeekam2023} shows that LLMs fail at strict logical reasoning tasks even when explicitly instructed. Previous work~\cite{Paper2} confirms this limitation in the IT-Grundschutz domain and motivates ASSERT's separation of LLM-based information extraction from deterministic graph verification via $\Delta G$.
\subsection{Question-Answering-Based vs. Graph-Level Verification}
To overcome the probabilistic limitations of text-based approaches, research is shifting toward KG-based compliance systems. The spectrum ranges from ontology-driven approaches such as the CO2 framework~\cite{Turaga2025}, which uses LLMs to extract a risk-based ontology from the EU AI Act, to dual-graph architectures such as GraphCompliance~\cite{GraphCompliance2025}, which aligns the policy and context graphs for GDPR scenarios, to multi-agent systems such as RAGulating Compliance~\cite{RAGulating2025} and AgCyRAG~\cite{AgCyRAG2025}. Further work includes the Neo4j-based Regulatory KG by Ershov~\cite{Ershov2023} and the ForPKG framework~\cite{ForPKG2025}. ForPKG is one of the few systems to quantify its extraction quality with precision (76.2\%) and recall (62.6\%).
Relevant contributions are emerging from related fields. CyberKG~\cite{CyberKG2025} uses a SecureBERT\_Plus-BiLSTM-Attention-CRF pipeline for cybersecurity KG construction and reports a macro-level entity-normalization F1 of 84.1\% for HAC-based clustering of synonymous CTI entities against human-annotated clusters, and the AEVS framework~\cite{AEVS2026} establishes an anchor-based verification method in which extracted triples are deterministically validated against source text anchors. The validation ensures the origin of the triples in the source text but does not check their topological correctness against a GT.
Despite this diversity, most systems do not evaluate the quality of the extracted graph itself. In particular, they rarely measure extraction quality using hard metrics such as precision, recall, and F1 score at the node or edge level. For example, CO2~\cite{Turaga2025} provides a qualitative proof of concept, RAGulating Compliance~\cite{RAGulating2025} reports no precision and recall for their triple extraction, Ershov~\cite{Ershov2023} demonstrates the graph via Cypher queries, and PrivComp-KG~\cite{PrivCompKG2024} uses only a global correctness score. AgCyRAG~\cite{AgCyRAG2025} similarly presents three qualitative use cases, builds a KG and then evaluates only the quality of downstream tasks such as question answering (Q\&A), without measuring the extraction quality of the graph itself. 
If a RAG agent answers a question correctly, the underlying graph is recorded as a ``success''. This represents a methodological weakness, as in regulated domains the graph is not just a heuristic search aid for RAG systems, but the factual basis upon which audits are based. The principle of accountability, as required by the BSI and NIS-2~\cite{NIS2}, demands that the verification process itself be auditable. A query result based on a faulty graph is of no regulatory value, as it hides topological flaws instead of revealing them.

While advanced systems such as ComplianceNLP~\cite{ComplianceNLP2026} rigorously evaluate the quality of regulatory semantic cross-references (F1 90.8\%), we are not aware of prior work that evaluates the structural correspondence of extracted IT-SC graphs against the company's GT. This structural correspondence involves the mathematical verification of the nodes and edges of an extracted infrastructure graph ($G_{Doc}$) against the validated infrastructure ($G_{GT}$). 
This differs  from the evaluation of semantic cross-references, as it does not concern linguistic references between paragraphs, but rather the correctness of actual infrastructure topologies. The ASSERT framework operationalizes this requirement through the deterministic graph comparator that calculates the graph difference $\Delta G = (V_O, V_P, E_O, E_T, E_G)$ as class-specific errors sets, thereby quantifying missing nodes, hallucinated nodes, and topologically incorrect edges.
\subsection{Positioning ASSERT}
ComplianceNLP~\cite{ComplianceNLP2026} operates exclusively within the forward paradigm for financial regulation and evaluates semantic cross-references, not infrastructure topologies. GraphCompliance~\cite{GraphCompliance2025} is GDPR-specific and validates graph quality via internal reconstruction stability (cycle consistency), in which a graph is iteratively converted to text and back, rather than by comparing it to an external infrastructure GT.
CO2~\cite{Turaga2025} evaluates structural graph characteristics (connectivity, density), but does not evaluate the accuracy of the extraction against a GT. 
Although AEVS~\cite{AEVS2026} follows a conceptually related anchor-based approach with deterministic restoration matching, it operates on a different ontological level. While AEVS validates extracted triples against the source text itself to ensure extraction accuracy, it primarily solves the linguistic NLP problem of making the graph a faithful representation of the potentially erroneous source. ASSERT, in contrast, addresses the broader audit challenge of verifying whether the constructed representation is aligned with an independently validated infrastructure reference graph ($G_{GT}$).
Since textual mappings do not imply factual accuracy, AEVS remains blind to substantive errors or outdated information within the legacy IT-SC itself. ASSERT overcomes this limitation using external mappings against the infrastructural GT. As a domain-independent framework, AEVS is not specialized for IT security. ASSERT thus combines all three aspects: reverse engineering of legacy IT-SC, deterministic graph-level verification, and schema-valid OSCAL outputs.
\section{Methodology: The ASSERT Framework}
\subsection{The ASSERT Framework Architecture}
ASSERT takes a legacy IT-SC and the infrastructure reference graph $G_{GT}$ from a preceding SA as input, constructs the document graph $G_{Doc}$, and verifies $G_{Doc}$ against $G_{GT}$, as shown in Fig.~\ref{fig:assert_architecture}. ASSERT treats the legacy IT-SC as potentially inconsistent and $G_{GT}$ as the trusted reference basis. It detects structural discrepancies after extraction, but does not validate incomplete or poisoned reference data or control effectiveness. The framework consists of three components:
\begin{figure}[htbp]
\centering
\scalebox{0.92}{%
\begin{tikzpicture}[
  font=\footnotesize,
  >={Stealth[length=2mm]},
  comp/.style={draw, thick, rounded corners=3pt, text width=6.2cm,
               inner sep=4pt, align=left, font=\footnotesize},
  io/.style={draw, fill=gray!15, rounded corners=2pt, text width=2.4cm,
             align=center, font=\footnotesize\bfseries, inner sep=2pt},
  gt/.style={draw, fill=green!15, rounded corners=2pt, text width=2.6cm,
             align=center, font=\footnotesize\bfseries, inner sep=2pt},
  hitl/.style={draw, dashed, fill=orange!25, rounded corners=2pt,
               text width=1.2cm, align=center,
               font=\scriptsize\bfseries, inner sep=2pt},
  arr/.style={->, thick},
  loop/.style={<->, thick, dashed, orange!70!black}
]
  % --- Three main components, vertically stacked ---
  \node[comp, fill=blue!8] (c1) at (0,0) {%
    \textbf{Information Extraction Component}
    \begin{list}{$\bullet$}{%
      \setlength{\topsep}{2pt}\setlength{\partopsep}{0pt}%
      \setlength{\itemsep}{0pt}\setlength{\parsep}{0pt}%
      \setlength{\leftmargin}{1.2em}\setlength{\labelwidth}{0.8em}%
      \setlength{\labelsep}{0.4em}\setlength{\rightmargin}{0pt}}%
      \item Hierarchical Provenance Chunking\\(Document $\to$ Chapter $\to$ Paragraph)
      \item LLM Extraction\\(Generic/Schema-Guided/Schema-Enforced)
      \item Deterministic Node Alignment $\to$ Exception List
    \end{list}
  };
  \node[comp, fill=violet!8, below=4mm of c1] (c2) {%
    \textbf{Dual-Graph Comparator ($\Delta G$)}\\[1pt]
    Stage~1 (Nodes): $V_O,\;V_P,\;V_{Matched}$\\
    Stage~2 (Edges, cascading): $E_O,\;E_T,\;E_G$
  };
  \node[comp, fill=green!10, below=4mm of c2] (c3) {%
    \textbf{OSCAL Export Module}\\[1pt]
    $G_{Verified}=(V_{Matched},\,E^M_{GT}\cap E^M_{Doc})$\\
    Filter $V_P,\,E_G,\,E_T$ $\to$ OSCAL SSP + AR
  };

  % --- Inputs / HITL (top) ---
  \node[io] (pdf)  at ([yshift=8mm,xshift=-22mm]c1.north)
        {Legacy\\IT-SC};
  \node[gt] (gt)   at ([yshift=8mm,xshift=6mm]c1.north)
        {$G_{GT}$\\from SA};
  \node[hitl] (hitl) at ([yshift=8mm,xshift=28mm]c1.north)
        {HITL\\Adj.};

  % --- Output (bottom) ---
  \node[io, below=4mm of c3] (out) {OSCAL SSP + AR};

  % --- Arrows: data flow ---
  \draw[arr] (pdf.south) -- (pdf.south |- c1.north);
  \draw[arr] (gt.south)  -- (gt.south  |- c1.north);
  \draw[arr] (c1) -- node[right=1pt, font=\scriptsize]{$G_{Doc}$} (c2);
  \draw[arr] (c2) -- node[right=1pt, font=\scriptsize]{$\Delta G$} (c3);
  \draw[arr] (c3) -- node[right=1pt, font=\scriptsize]{$G_{Verified}$} (out);

  % --- HITL bidirectional loop (Exception List <-> Auditor) ---
  \draw[loop] (hitl.south) -- (hitl.south |- c1.north);
\end{tikzpicture}%
}
\caption{Architecture of the ASSERT framework for transforming legacy IT-SC into schema-valid OSCAL artifacts. Dashed lines mark the optional HITL review, and $G_{GT}$ is the verified infrastructure GT of the SA from~\cite{Paper2}.}
\label{fig:assert_architecture}
\end{figure}
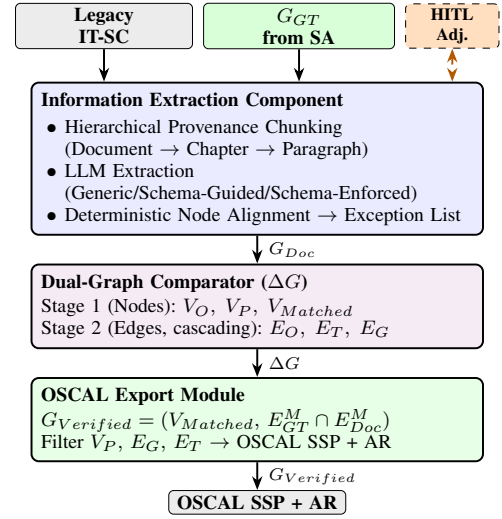
\newline 
\textbf{1. The Information Extraction Component:}
This component constructs $G_{Doc}$ from the legacy IT-SC by extracting entities and relationships under the selected ontology exposure level. Since Grundschutz-based IT-SCs follow strict chapter guidelines, ASSERT uses this existing structure for ontology-based information extraction.
\newline
\textit{1. Hierarchical Chunking \& Extraction:} To ensure the traceability of each extracted entity, the legacy IT-SC is modeled as a hierarchical tree when ingested into the graph. Chunks contain on average $\approx 3,700$ tokens and preserve document structure. Oversized chapters are split at sections, tables, or paragraphs, without sliding-window overlap. An LLM extraction step operates the chunks at the selected ontology exposure level. The extraction validates the LLM response against a Pydantic schema. In Schema-Enforced mode, a deterministic post-hoc step filters against the canonical entity list from $G_{GT}$.
\newline
\textit{2. Automated Node Alignment:} The node alignment between $G_{Doc}$ and $G_{GT}$ follows a cascading, strictly lexical procedure to rule out stochastic misalignments and ensure the reproducibility of the difference calculation in contrast to HybridRAG approaches~\cite{Sarmah2024, Paper2}. A match is defined in three priority levels: case-insensitive identity of the identifiers (1), identity of the normalized abbreviations (2), mutual substring inclusion of name or alias (3). This assumes the naming consistency expected in BSI IT-Grundschutz documents derived from structured asset tables. Matches that are successfully identified are deduplicated and added to the set of matching nodes $V_{Matched}$ used by the comparator. Mentions that do not match at any of these levels are moved to the exception list $V_{Exc}$ and reviewed by a Human-in-the-Loop (HITL).         
\newline
\textit{3. Human-in-the-Loop Review:} 
The final decision regarding the entities on the exception list is delegated to the HITL. Since $V_{Exc}$ contains all unresolved mentions before review, it quantifies the mention-level HITL workload of a given run. For each ambiguous entity, the expert is shown the original text fragment. The expert can verify the context and decide to assign the entity to an existing asset or to mark it as incorrect or new. 
Consequently, the exception list addresses ontological 1:N mismatches between the text and $G_{GT}$, which the HITL resolves before they are incorrectly reported as $\Delta G$.
\newline
\textbf{2. The Dual-Graph Comparator:}
In this component, $G_{Doc}$ and $G_{GT}$ are compared algorithmically. In accordance with the defined cascading logic, the comparator first determines the sets $V_O$, $V_P$, and $V_{Matched}$ at the node level through deterministic string matching against $G_{GT}$. It then calculates, at the edge level (conditional on $V_{Matched}$), the three edge classes of the compliance difference $\Delta G = (V_O, V_P, E_O, E_T, E_G)$ over directed $(u,v)$-tuples: missing edge tuples ($E_O$) via the difference $E_{GT} \setminus E_{Doc}$ over $V_{Matched}$, unsupported document edge tuples ($E_T$) via the difference $E_{Doc} \setminus E_{GT}$ over $V_{Matched}$, and ghost edges ($E_G$) by filtering all edges with at least one phantom node from $V_P$.
\newline
\textbf{3. The OSCAL Export Module:} 
ASSERT separates system description from verification evidence. The verified component set $V_{Matched}$ and BSI Grundschutz++ control implementation statements are assembled into an SSP, while an AR records graph-difference findings and source-linked evidence. Because the published BSI Grundschutz++ OSCAL artifacts provide the requirement catalog but no profile for the claimed control subset, ASSERT uses a minimal local profile and Assessment Plan to connect the catalog, SSP, and AR.
\section{Dataset and Experimental Setup}
The ASSERT framework is evaluated on the RecPlast dataset using three configurations with different levels of reference-ontology exposure: Generic, Schema-Guided, and Schema-Enforced. The ASSERT pipeline uses LangChain for orchestration with Pydantic-typed outputs. To investigate the model dependency of the results, inference is performed using Ollama with the local open-weight model Gemma~4 26B and the commercial model Anthropic Claude Opus~4.7. The generated SSP and AR are validated against the OSCAL v1.1.3 schemas.
\subsection{Dataset - RecPlast GmbH}
The evaluation of ASSERT requires a public dataset from which a validated reference graph can be reconstructed. Since real-world IT-SCs are generally not publicly available due to non-disclosure agreements (NDAs), this study uses the expert-generated ``RecPlast GmbH'' dataset~\cite{BSIRecPlast} published by the BSI. The dataset covers the full IT-Grundschutz certification process chain, including the initial organizational and infrastructure descriptions, the intermediate SA artifacts, the final 69-page IT-SC, and more. This allows us to reconstruct $G_{GT}$ from the intermediate artifacts, construct $G_{Doc}$ independently from the final IT-SC, and compare both graphs under reproducible conditions.
\subsection{Evaluation Metrics}
Precision, recall, and F1 score are computed at the node and edge levels and for specific fault classes. In the baseline setting, $\Delta G$ denotes the diagnostic delta sets produced by the graph comparator, and class-specific detection scores are defined only for the fault-injected variants.

\textbf{Node-level metrics:} The vertex sets are evaluated at the entity level. $V_{Matched}$ counts the unique GT nodes that have received at least one aligned extraction, and $V_P$ counts the unique phantom entities after deduplication via (name, type). $V_{Doc}$ stays at mention level and reflects extraction and HITL workload. This results in $P_V = |V_{Matched}|/(|V_{Matched}|+|V_P|)$, and $R_V = |V_{Matched}|/|V_{GT}| = |V_{Matched}|/(|V_{Matched}|+|V_O|)$. $F_{1,V}$ is defined as the harmonic mean of $P_V$ and $R_V$.

\textbf{Edge-level metrics:} Edge evaluation is restricted to the subgraphs induced by $V_{Matched}$ to avoid penalizing missing nodes twice. As defined above, the edge error classes $E_O$, $E_T$, and $E_G$ use deduplicated directed endpoint tuples $(u,v)$. For edge-quality metrics, relation-label agreement is additionally required. Let $E^M_{GT,r}$ and $E^M_{Doc,r}$ denote the matched-node-induced edge sets over deduplicated directed triples $(u,r,v)$, where $r$ is the stored relation type after case and whitespace normalization and malformed or out-of-schema labels remain nonmatching. Consequently, $E_{Match,r}=E^M_{Doc,r}\cap E^M_{GT,r}$, $P_{E,r}=|E_{Match,r}|/|E^M_{Doc,r}|$, $R_{E,r}=|E_{Match,r}|/|E^M_{GT,r}|$, and $F_{1,E,r}$ is their harmonic mean.

\textbf{Class-specific metrics:} F1 scores are calculated for the five error classes ($V_O, V_P, E_O, E_T, E_G$) in the fault injection evaluation, as shown in Table~\ref{tab:fault_consolidated}, where the $k=20$ injected errors define the GT. Whether a unit is considered a match depends on the error class: $V_O$ is matched via infrastructure node IDs, $V_P$ via normalized entity names, and edge errors ($E_O$, $E_T$, $E_G$) via directed endpoint pairs. As false positives, only additional detections compared to the unmodified RecPlast IT-SC (baseline) are counted. Such class-specific metrics are not defined on the baseline IT-SC because the actual distribution of inconsistencies is not known. The baseline counts in Table~\ref{tab:ablation} are therefore reported separately from class-specific P/R/F1 values. Baseline edge-error counts use endpoint tuples, and edge metrics use typed triples.

\subsection{ASSERT Extraction Configurations}
To assess how ontology exposure affects information extraction, the ASSERT framework is evaluated in three configurations. These are defined as follows:

\textbf{ASSERT Generic:} The LLM extracts entities and relationships solely on the basis of the seven abstract entity classes from the BSI SA (IT system, application, business process, room, communication link, IoT system, ICS system), without any knowledge of the specific component names or dependencies within the company. The LLM thus recognizes the categories but not the instances of the infrastructure. This configuration serves as a methodological baseline, as it isolates the LLM's document-based extraction capability and quantifies the effect of reference-ontology exposure.

\textbf{ASSERT Schema-Guided:} The LLM is provided with the complete ontology of the reference graph $G_{GT}$, including all specific entity names with their types and abbreviations, as well as all known relations with their dependency types. The LLM is instructed to use the known component names exactly as they appear in the text, but is also permitted to extract entities that are not listed in the ontology. The comparison with Generic quantifies how reference-ontology exposure affects recall and precision while preserving non-reference extraction.

\textbf{ASSERT Schema-Enforced:} The schema is strictly enforced on two levels. First, the prompt instructs the LLM to extract only entity and relation types defined in the $G_{GT}$ schema. Second, a deterministic post-hoc filter removes all extracted entities whose names do not appear in the canonical list. This dual strategy eliminates non-canonical phantom nodes ($V_P=0$) by construction. However, this simultaneously suppresses the detection of entities that exist in the real infrastructure but are not modeled in $G_{GT}$, e.g., shadow IT. 
For all ASSERT configurations, $\Delta G$ is computed with the same deterministic comparator against $G_{GT}$, ensuring metric comparability across all configurations. To quantify model dependence, all configurations are evaluated with both LLMs.
\subsection{Fault-Injected Classes ($V_O, V_P, E_O, E_T, E_G$)}
The fault-injected evaluation allows an isolated assessment of the framework's class-specific sensitivity. Therefore, the baseline IT-SC is selectively altered, and the framework is run against the modified IT-SCs. In the single-fault variants, only one error class is introduced per run, so that the measured detection rates can be causally attributed to that specific class. The writing style, terminology, and contextual structure of the audit report remain unchanged.
Specifically, for $V_O$, $k=20$ entities and their text contexts are removed from the baseline IT-SC. For $V_P$, we inject fictional entities from a curated list, e.g., the mention of a non-existent ``cloud backup''. For $E_G$, we inject relations incident to phantom entities, such as an alleged connection from the fictional entity to the production network. For $E_O$, we delete sentences that explicitly name edges, with the constraint that both endpoints co-occur in the same sentence and both endpoints fall within the entity intersection of all three ASSERT configurations. For $E_T$, the causality of $k=20$ edges explicitly named in the text is manipulated. $k=20$ is the largest feasible RecPlast intersection size, covering approximately 15\% of $G_{GT}$, creating substantial but not syntactically destructive document alteration.
The mixed variant ($V_{mixed}$) injects $k'=5$ errors per class, resulting in 25 simultaneous manipulations to test robustness under coexisting error types.
\section{Results \& Discussion}
The baseline IT-SC compares the three configurations across both LLMs and quantifies the resulting node- and edge-level graph differences, as shown in Table~\ref{tab:ablation}. The fault-injected variants isolate the framework's sensitivity to specific error classes under controlled document manipulations, as shown in Table~\ref{tab:fault_consolidated}.
\begin{table}[htbp]
\caption{Extraction metrics for ASSERT configurations.}
\label{tab:ablation}
\centering
\scriptsize
\begin{tabular*}{\columnwidth}{@{\extracolsep{\fill}}lcccccc@{}}
\hline
 & \multicolumn{3}{c}{\textbf{Gemma 4}} & \multicolumn{3}{c}{\textbf{Opus 4.7}} \\
  \cline{2-7}
   \textbf{Class}&  Generic & Guided & Enforced & Generic & Guided & Enforced  \\

\hline
$V_{Doc}$      & 321 & 586 & 222 & 123 & 126 & 134 \\
$V_{Matched}$  & 95  & 132 & 123 & 120 & 118 & 130 \\
$V_{Exc}$    & 134 & 182 & 0   & 0   & 4   & 0   \\
$V_O$          & 39  & 2   & 11  & 14  & 16  & 4   \\
$V_P$          & 133 & 172 & 0   & 0   & 4   & 0   \\
$E_O$          & 469 & 637 & 731 & 645 & 621 & 732 \\
$E_T$          & 16  & 33  & 4   & 0   & 22  & 0   \\
$E_G$          & 35  & 49  & 0   & 0   & 15  & 0   \\
\hline
$P_V$   & 0.417 & 0.434 & 1.000 & 1.000 & 0.967 & 1.000 \\
$R_V$   & 0.709 & 0.985 & 0.918 & 0.896 & 0.881 & 0.970 \\
$F_{1,V}$  & 0.525 & 0.603 & 0.957 & 0.945 & 0.922 & 0.985 \\
\hline
$P_{E,r}$      & 0.133 & 0.843 & 0.931 & 0.891 & 0.687 & 0.895 \\
$R_{E,r}$      & 0.008 & 0.261 & 0.069 & 0.147 & 0.137 & 0.137 \\
$F_{1,E,r}$    & 0.016 & 0.399 & 0.128 & 0.253 & 0.229 & 0.238 \\
\hline
\multicolumn{7}{@{}p{0.96\columnwidth}@{}}{Results for Gemma 4 and Opus 4.7 across ASSERT configurations, reporting node counts, endpoint-tuple edge-error counts, and typed-triple edge metrics.}
\end{tabular*}
\end{table}
\begin{figure}[t]
\centering
\scriptsize
\setlength{\tabcolsep}{2.2pt}
\renewcommand{\arraystretch}{1.35}
\begin{tabular}{|c|c|c|c|}
\hline
\rowcolor{headblue}
\parbox[c][0.48cm][c]{1.20cm}{\centering} &
\parbox[c][0.48cm][c]{2.20cm}{\centering\textbf{Generic}} &
\parbox[c][0.48cm][c]{2.20cm}{\centering\textbf{Guided}} &
\parbox[c][0.48cm][c]{2.20cm}{\centering\textbf{Enforced}} \\
\hline

\rowcolor{useblue}
\parbox[c][0.58cm][c]{1.20cm}{\centering\textbf{Best use}} &
\parbox[c][0.58cm][c]{2.20cm}{\centering Discovery\\baseline} &
\parbox[c][0.58cm][c]{2.20cm}{\centering Balanced\\analysis} &
\parbox[c][0.58cm][c]{2.20cm}{\centering Phantom-free\\OSCAL export} \\
\hline

\rowcolor{strengthgreen}
\parbox[c][0.78cm][c]{1.20cm}{\centering\textbf{Strength}} &
\parbox[c][0.78cm][c]{2.20cm}{\centering Finds non-\\reference assets} &
\parbox[c][0.78cm][c]{2.20cm}{\centering Improves edge/\\conflict extraction} &
\parbox[c][0.78cm][c]{2.20cm}{\centering Removes phantom\\entities by design} \\
\hline

\rowcolor{problemred}
\parbox[c][0.68cm][c]{1.20cm}{\centering\textbf{Problem}} &
\parbox[c][0.68cm][c]{2.20cm}{\centering Noisy output;\\high HITL burden} &
\parbox[c][0.68cm][c]{2.20cm}{\centering Model-dependent;\\ontology leakage} &
\parbox[c][0.68cm][c]{2.20cm}{\centering No discovery;\\masks shadow IT} \\
\hline

\rowcolor{edgeamber}
\parbox[c][0.58cm][c]{1.20cm}{\centering\textbf{Edge}} &
\parbox[c][0.58cm][c]{2.20cm}{\centering Many edges\\remain missing} &
\parbox[c][0.58cm][c]{2.20cm}{\centering Best edge\\trade-off} &
\parbox[c][0.58cm][c]{2.20cm}{\centering Clean nodes,\\low edge recall} \\
\hline

\rowcolor{decisiongray}
\parbox[c][0.58cm][c]{1.20cm}{\centering\textbf{Use when}} &
\parbox[c][0.58cm][c]{2.20cm}{\centering Unknown assets\\matter} &
\parbox[c][0.58cm][c]{2.20cm}{\centering Model fidelity\\is reliable} &
\parbox[c][0.58cm][c]{2.20cm}{\centering Export validity\\matters} \\
\hline
\end{tabular}
\caption{Evaluation-derived comparison of ASSERT extraction modes.}
\label{fig:mode_comparison}
\end{figure}
\newline
\textbf{ASSERT Configurations:}
Table~\ref{tab:ablation} and Fig.~\ref{fig:mode_comparison} show that the three configurations serve different operational roles. Generic preserves discovery, but creates substantial HITL workload for the Gemma model. Schema-Guided improves edge-related extraction, but the benefit is model-dependent. For Gemma, Schema-Guided reduces $V_O$ from 39 to 2 and increases recall ($R_V=0.709 \to 0.985$), but this comes at the cost of massive over-extraction, as $V_{Doc}$ increases from 321 to 586 and $V_P$ from 133 to 172, leaving precision low ($P_V=0.434$). This indicates stronger alignment with $G_{GT}$ rather than document-only extraction. The Opus model only shows marginal over-extraction ($V_P=4$ in Schema-Guided, $V_P=0$ otherwise) and maintains $P_V\geq0.967$ across configurations, suggesting that the phantom problem is primarily model-dependent instruction fidelity.
Schema-Enforced achieves high node-level $F_{1,V}=0.957$ for Gemma and $F_{1,V}=0.985$ for Opus and removes phantom nodes and ghost edges by construction, making it suitable for an OSCAL export. However, $V_P=0$ is ambiguous, as it eliminates both hallucinations and discovery capability. A shadow IT asset absent in $G_{GT}$ is systematically ignored, since the model can only emit canonical-list entities. The high $E_O$ counts complement this. A typical IT-SC states ``Server~A runs App~B'', but does not mention every dependency recorded in the tabular SA. Despite high node performance, typed edge recall remains low across all configurations ($R_{E,r}\leq0.261$) because the IT-SC identifies only a subset of SA dependencies. Typed-edge performance therefore remains model-dependent, and unrestricted extraction can generate unverifiable relations.
\newline
\textbf{Pipeline Accuracy under Fault Injection:}
Table~\ref{tab:fault_consolidated} reports class-specific F1 scores for single-fault and mixed variants, isolating extraction accuracy from document completeness.
\begin{table}[htbp]
\caption{Fault-injected evaluation.}
\label{tab:fault_consolidated}
\centering
\footnotesize
\setlength{\tabcolsep}{4pt}
\begin{tabular}{@{}lc ccc ccc@{}}
\hline
 & & \multicolumn{3}{c}{\textbf{Gemma 4}} & \multicolumn{3}{c}{\textbf{Opus 4.7}} \\
\cline{3-8}
\textbf{Class} & $\textbf{k}$ & Generic & Guided & Enforced & Generic & Guided & Enforced  \\
\hline
\multicolumn{8}{@{}l@{}}{\textit{Single-Fault Classes}}            \\
$V_O$       & 20 & 0.667 & 0.258 & 0.556 & 0.645 & 0.597 & 0.526   \\
$V_P$       & 20 & 0.488 & 0.741 & 0.000    & 0.889 & 1.000 & 0.000      \\
$E_G$       & 20 & 0.526 & 0.630 & 0.000    & 0.884 & 1.000 & 0.000      \\
$E_O$       & 20 & 0.109 & 0.200 & 0.082 & 0.200 & 0.246 & 0.000      \\
$E_T$       & 20 & 0.000    & 0.303 & 0.000    & 0.086 & 0.710 & 0.102   \\
\hline
\multicolumn{8}{@{}l@{}}{$V_{mixed}$ \textit{(simultaneous injection of multiple classes)}} \\
$V_O$       &  5 & 0.556 & 0.049 & 0.000      & 0.417 & 0.333 & 0.074 \\
$V_P$       &  5 & 0.137 & 0.066 & 0.000      & 1.000 & 1.000 & 0.000    \\
$E_G$       &  5 & 0.118 & 0.200 & 0.000      & 0.286 & 1.000 & 0.000    \\
$E_O$       &  5 & 0.029 & 0.049 & 0.061 & 0.068 & 0.080 & 0.000      \\
$E_T$       &  5 & 0.167 & 0.000 & 0.250 & 0.000      & 0.750 & 0.000    \\
\hline
\multicolumn{8}{@{}p{0.96\columnwidth}@{}}{\scriptsize Rows denote injected error classes. Columns show class-specific F1 scores for (Generic / Schema-Guided / Schema-Enforced) under Gemma 4 and Opus 4.7. $V_{mixed}$ uses simultaneous $k'=5$ injections per class.}
\end{tabular}
\end{table}
Generic reliably captures phantom classes if the model is obedient to instructions. Opus achieves $F_1=0.889$/$0.884$ for $V_P$/$E_G$, while Gemma falls behind with $F_1=0.488$/$0.526$ due to lower instruction adherence. By design, Enforced achieves $F_1=0$ for $V_P$ and $E_G$. By definition, the dual strategy of prompt constraints and post-hoc filtering prevents phantom detection. $E_T$ remains the most challenging error class, with a substantial model gap. In the Schema-Guided configuration, Opus 4.7 achieves $F_1=0.710$, while Gemma 4 lags behind with $F_1=0.303$. This twofold F1 gap quantifies the trade-off between data sovereignty (local LLM) and detection capability on dense SA tables. Even with Opus, topological-conflict detection remains incomplete. ASSERT does not resolve the underlying reasoning limitation, but makes its occurrence auditable through deterministic graph matching.
Schema-Guided has opposite effects on the two LLMs. For Opus, it is the most balanced configuration, with perfect phantom detection, $V_O$ is slightly below Generic at $F_1=0.597$, $E_O$ increases slightly ($0.200 \to 0.246$), while $E_T$ benefits massively ($0.086 \to 0.710$). Conversely, Gemma Schema-Guided reveals empirically verifiable ontology-induced hallucination, as 16 of 20 omissions go unnoticed. The $V_O$ F1 score drops from $0.667$ (Generic) to $0.258$, and recall drops from $0.700$ to $0.200$. A manual analysis showed that, for 12 of the 16 unrecognized $V_O$ omissions, Gemma hallucinates the removed node back into the output under reference-ontology exposure.
Once the document deviates from the reference ontology $G_{GT}$, the trade-off not observed on the baseline IT-SC emerges. For Gemma, the $V_O$ F1 score drops from $0.667$ (Generic) to $0.556$ (Schema-Enforced). Recall increases slightly ($0.750$ vs. $0.700$), but precision drops from $0.636$ to $0.441$ because the rigid lexical filter generates pseudo-omissions for textually divergent identifiers. For $E_O$, Enforced degrades significantly (Gemma recall $R=0.200$, Opus recall $R=0.000$) because the canonical filter classifies the $E_O$ targets in the entity intersection as $FN$. The convergence of both LLMs in Enforced configuration on the baseline IT-SC is therefore a RecPlast artifact, as the LLMs diverge again due to wording differences.
The $V_{mixed}$ variant exhibits a robustness asymmetry. Opus maintains stable $E_T$ detection under simultaneous manipulations ($F_1=0.750$ vs. $0.710$ for single-fault), while Gemma fails ($F_1=0.000$ vs. $0.303$). The breakdown manifests in schema violations such as hallucinated relation types ($144$ of $183$ relations with the literal value ``relation\_type''), phantom mass extraction ($\textit{FP}=142$ for $V_P$), and typos such as ``depends\_onn''. Multiple simultaneous error classes appear to undermine the LLM's instruction fidelity, leading to a breach of schema conformity. This limits the suitability of fully local deployments for CI audits with complex dependency structures and shows that single-fault conditions underestimate the robustness gap between local and cloud-based LLMs. Beyond its diagnostic value, the Gemma $V_{mixed}$ Schema-Guided run provides a controlled stress case for the HITL interface, yielding the largest exception list observed in our experiments ($V_{Exc}=207$ mentions).
The results narrow down the mode selection to a trade-off between discovery and reference control. Generic retains discovery capability but requires HITL to distinguish genuine new discoveries from hallucinations. Schema-Guided is most balanced when instruction obedience is high (Opus), but can intensify ontology-induced hallucinations when it is low (Gemma). Schema-Enforced eliminates phantom entities by design ($V_P=0$, $E_G=0$), but excludes discovery and degrades $V_O$ and $E_O$ under wording deviation. Thus, the configuration selection mainly depends on model instruction adherence and the desired degree of discovery versus reference control.
\newline
\textbf{OSCAL Output Module:}
As a proof of concept, Schema-Enforced outputs were exported into OSCAL SSP and AR artifacts. The SSPs contain 222 (Gemma) and 134 (Opus) components at the mention level, corresponding to 123 and 130 unique GT entities, while the ARs include graph-difference findings and source-linked evidence. With $V_P=0$, the SSP export is free of non-canonical phantom entities but is not discovery-capable.
\section{Limitations \& Future Work}
Naturally, ASSERT remains dependent on the quality of its input artifacts. Undocumented assets and shadow IT cannot be extracted or flagged as missing. Errors or inconsistencies in $G_{GT}$ propagate directly into $\Delta G$. ASSERT therefore shifts part of the verification burden to establishing and maintaining a reliable reference graph. Reference-ontology exposure can additionally mask document deviations when models reconstruct missing document evidence from $G_{GT}$. Future work should investigate industry-specific expectation ontologies and methods for validating $G_{GT}$ across multiple operational sources, confidence scores, and alternative comparison strategies.
Further, the evaluation is limited by the RecPlast dataset and the chosen fault injection method. Typed edge metrics are likewise limited to relation labels represented in the reconstructed reference graph. RecPlast is the only fully public, expert-generated IT-Grundschutz dataset, but it does not allow for broad generalizations to diverse real-world IT-SCs. The fault injection size ($k=20$, $k'=5$) is constrained by the entity overlap in RecPlast. Since runs were not repeated across seeds, the results should be interpreted as a proof-of-concept evaluation rather than a statistically conclusive benchmark.
From a practical perspective, the Schema-Enforced configuration is only suitable as a complementary filter, since it prevents phantom entities but may mask real infrastructure not included in $G_{GT}$. Future work should include robust multi-fault evaluations, schema-following tuning, and cascaded multi-pass validation for local models, and neuro-symbolic methods for the deterministic resolution of remaining $E_T$ conflicts~\cite{Hsia2025}. The OSCAL export should be expanded from a schema-valid artifact to a semantic cross-catalog alignment between IT-Grundschutz and Grundschutz++.
While ASSERT is currently aligned with BSI IT-Grundschutz, OSCAL suggests broader applicability beyond German regulatory scenarios. Future work should investigate adaptations to other OSCAL-representable control catalogs, such as NIST SP 800-53 or ISO/IEC 27001.
%While ASSERT is currently aligned with BSI IT-Grundschutz, its use of OSCAL suggests a path toward broader applicability beyond German regulatory scenarios. Future work should investigate adaptations to other OSCAL-representable control catalogs, such as NIST SP 800-53 or ISO/IEC 27001, by extending the SA agent ontology.
%
The exception list ranges from $0$ items (Opus Generic/Enforced) to $182$ on the baseline RecPlast IT-SC and reaches $207$ in the Gemma Schema-Guided stress case. These counts quantify the potential HITL workload, while actual review time and cognitive load remain subject to controlled user evaluation.
\section{Conclusion}
ASSERT addresses the migration of text-based legacy IT-SCs to machine-readable compliance artifacts by separating graph-verified content from documented inconsistencies before export. During the multi-year transition period between IT-Grundschutz and Grundschutz++, it ensures consistency between narrative IT-SC and expert-verified infrastructure reference graphs $G_{GT}$. Formal document graphs ($G_{Doc}$) are constructed from IT-SCs under different levels of reference-ontology exposure, deterministically aligned with $G_{GT}$, and discrepancies are made auditable as graph differences ($\Delta G$).
The RecPlast evaluation shows that ASSERT detects topological conflicts against the reconstructed reference graph, with model-dependent performance. The fault-injected evaluation highlights the trade-off between cloud-based LLM quality and on-premises data sovereignty, as well as a robustness asymmetry among coexisting error classes.
The schema-valid OSCAL SSP and AR outputs separate verified system content from verification evidence and demonstrate syntactic compatibility with OSCAL-based CaC workflows. ASSERT thereby provides a deterministic comparison and migration mechanism for ontology-based compliance graphs in regulated audit scenarios.
Beyond detection quality, the exception-list provides an observable proxy for potential HITL review burden, while actual review time and cognitive load remain to be evaluated.

\end{document}